# Internal transitions of quasi-2D charged magneto-excitons in the presence of purposely introduced weak lateral potential energy variations

C. J. Meining, V. R. Whiteside, B. D. McCombe

*Department of Physics and CSEQuIN, University at Buffalo, SUNY, Buffalo, NY 14260, USA*

A. B. Dzyubenko[*]

*Department of Physics, CSU at Bakersfield, CA 93311, USA*

J. G. Tischler, A. S. Bracker, D. Gammon

*Naval Research Laboratory, Washington, D.C. 20375-5347, USA*

## Abstract

Optically detected resonance spectroscopy has been used to investigate effects of weak random lateral potential energy fluctuations on internal transitions of charged magneto-excitons (trions) in quasi two-dimensional GaAs/AlGaAs quantum-well (QW) structures. Resonant changes in the ensemble photoluminescence induced by far-infrared radiation were studied as a function of magnetic field for samples having: 1) no growth interrupts (short range well-width fluctuations), and 2) intentional growth interrupts (long range monolayer well-width differences). Only bound-to-continuum internal transitions of the negatively charged trion are observed for samples of type 1. In contrast, a feature on the high field (low energy) side of electron cyclotron resonance is seen for samples of type 2 with well widths of 14.1 and 8.4 nm. This feature is attributed to a bound-to-bound transition of the spin-triplet with non-zero oscillator strength resulting from breaking of translational symmetry.

Internal transitions of neutral and charged excitons depend strongly on the symmetry properties of these electron-hole complexes in quasi-two-dimensional (2D) GaAs/AlGaAs quantum wells (QWs) in an applied magnetic field. For example, for the neutral quasi-2D exciton (X), the axial symmetry of the electron-hole pair in a magnetic field results in a direct relationship between the energies of excited, low-field p-like states of the X and electron/hole single particle cyclotron energies. This result is independent of magnetic field and holds even for the complex valence band Landau level (LL) structure of a QW, where heavy-hole and light-hole levels strongly interact provided the warping terms are weak.[1,2] With two electrons bound to a single hole, a negatively charged exciton (X⁻), a symmetry related to the center-of-mass (CM) motion leads to surprisingly strict selection rules that are apparent in magneto-optics.[3,4] In contrast to the superficially similar negatively charged donor D⁻,[5-8] the positive charge of an X⁻-complex is free to move in the QW plane. Consequently, internal bound-to-bound transitions are forbidden, and only bound-to-continuum transitions are allowed.[3,4] When the magnetic translational symmetry of the X⁻-complex is removed, however, the strict selection rules are broken, and bound-to-bound transitions acquire a non-zero oscillator strength.[3,9]

We report effects of purposely introduced, weak, random lateral potential energy variations (due to monolayer (ML) well-width fluctuations) on internal transitions of X⁻. The consequences of symmetry breaking by these potential energy variations, whose characteristic length scale is greater than the exciton Bohr radius, are revealed directly by optically detected resonance (ODR) spectroscopy. In this highly sensitive technique[6] internal transitions from the ground state of X⁻ associated with the lowest electron Landau level ($n_e = 0$) to excited states associated with the first excited Landau level ($n_e = 1$) are studied by monitoring resonant changes of the (ensemble) photoluminescence (PL) induced by absorption of a far-infrared (FIR) laser beam in a magnetic field. We compare data for: 1) a QW with no growth interrupts (lateral range of well-width fluctuations is small on the scale of the exciton Bohr radius), and 2) a QW for which lateral fluctuations of the well-width with length scales of several Bohr radii have been purposely introduced by interrupting the epitaxial growth at the interface. These studies focus on samples with very weak interface potential energy fluctuations in relatively wide wells, which do not confine excitons at the temperatures of the experiment; the corresponding energy fluctuation in narrow wells are large enough to weakly confine excitons in interface-fluctuation quantum dots, IFQDs.[10, 11] These studies, which are distinct from recent ODR investigations of ensembles of strongly confining self-assembled InAs/GaAs[12] and InSb/GaSb[13] quantum dots, provide insight into various mechanisms (currently under debate) that lead to breaking of translational invariance.

Two GaAs/AlGaAs QW samples grown by molecular beam epitaxy on semi-insulating GaAs (100) substrates were studied. Sample 1 (no growth interrupts) consists of 40 repetitions of 20 nm GaAs wells sandwiched between 40 nm $Al_{0.3}Ga_{0.7}As$ barriers, δ-doped in the barrier centers with silicon at $2\times10^{10}$ cm$^{-2}$. In such samples the characteristic lateral length scale of interfacial roughness is much smaller than the exciton Bohr diameter.[14, 15] Sample 2 contains five single QWs of nominal widths 2.8, 4.2, 6.2, 8.4 and 14.1 nm separated by 40 nm $Al_{0.3}Ga_{0.7}As$ barriers with the narrowest well closest to the sample surface.[16] The barriers that were grown after each well were doped with Si at $10^{17}$ cm$^{-3}$ over a 3 nm region starting 10 nm after completing growth of the well; the resulting sheet electron density in each well is approximately $3\times10^{10}$ cm$^{-2}$, close to that of sample 1. The growth was interrupted at the well/barrier interfaces leading to formation of monolayer high lateral regions that serve as natural quantum dots[17] for the narrower QWs (2.8 and 4.2 nm widths). Lateral localization of neutral and charged excitons has been demonstrated for the two narrowest wells in these samples by the very sharp homogeneously broadened PL lines[18] obtained from single dot studies. Here, we concentrate on results from the widest well (14.1 nm); the 8.4 nm QW shows qualitatively and quantitatively very similar results, which are not shown here in the interest of brevity. The monolayer fluctuations lead to small (< 1 meV) lateral potential energy variations with characteristic lateral length scale between 20 and 40 nm. Nevertheless, these weak lateral fluctuations in potential energy are manifest through effects on the internal transitions, since the characteristic lateral length scale is larger than the exciton Bohr radius. The ODR spectra for the narrowest wells (not shown here), on the other hand, can be understood as an inhomogeneously broadened ensemble of internal transitions of negatively charged excitons confined in an array of quantum dots having a range of lateral dimensions.[11]

ODR spectra for sample 1 were obtained in a 9 T superconducting magnet system in conjunction with a previously described fiber optic/far-infrared light-pipe setup.[2, 6] Measurements on sample 2 were taken with focusing optics for both visible (HeNe laser) and far IR beams (from an Edinburgh Instruments Model FIRL100 optically pumped laser).[11] Magnetic fields were provided by an Oxford Instruments 10T optical-access superconducting magnet system with a variable temperature insert. All data were taken in the Faraday geometry at 4.2 K.

Photoluminescence spectra of both samples show multiple features that have been previously observed and identified[16, 19] as recombination from neutral and negatively charged excitons. In the case of sample 2 additional "fine" structure is observed. The measured separation of corresponding features (X or X$^-$) due to one ML difference in well width (MLS) is approximately 0.75 meV for the 14.1 nm QW, in reasonable agreement with calculations of the lowest QW subband energies of finite barrier GaAs/Al$_{0.3}$Ga$_{0.7}$As QWs with typical parameters.[20]

The zero field trion binding energies for sample 1 and the 14.1 nm well of sample 2 are about 1.3 meV and 1.4 meV, respectively.

Figure 1(c) shows the PL at 6.24 T (CCD detection), the field corresponding to electron cyclotron resonance (eCR), with resolution (0.8 meV) insufficient to resolve the MLS. The PL is shown without (dashed line) and with (solid line) simultaneous FIR illumination (P = 180 mW) at $E_{FIR}$ = 10.43 meV. At this FIR photon energy and magnetic field, free electrons in the well undergo cyclotron resonance transitions between the two lowest LLs and resonantly heat the carrier system leading to an increase of the X population and a decrease of the $X^-$ population. This gives rise to a negative ODR signal (black hatched region) in Fig. 1(c) for the latter and a positive ODR signal (white hatched region) for the former. This process involves a combination of partial ionization of the photo-excited $X^-$-complexes and electron redistribution in the laterally modulated potential energy landscape (indicated schematically in Fig. 1(d)). Figure 1(a) shows a grayscale contour plot of the ODR signal strength as a function of PL energy and magnetic field ($\Delta B = 0.02$ T). Dark (bright) contours correspond to negative (positive) ODR signal, respectively. The dotted vertical line marks eCR, for which a constant field ODR signal (the difference between FIR laser on and off) is shown by the dotted white line in Fig. 1(c). Note that the $X^-$ (X) ODR signal remains negative (positive) over the entire field range, and that both features shift to higher energies with increasing field in accord with the diamagnetic shift of the excitonic complexes. The field dependence of the ODR signal can be obtained by tracking one of the PL features. The two black lines in Fig. 1(a) indicate the energy bandpass (0.6 meV) over which the ODR signal for the $X^-$ feature was averaged to obtain the thick black line of Fig. 1(b). Similar results (thin white line in Fig. 1(b)) can be obtained with a chopped FIR beam, a photomultiplier detector, and a lock-in amplifier, while tracking $X^-$ by programming the spectrometer drive to follow its diamagnetic shift; the spectral window is set by the spectrometer slits. Inverted signals are obtained when tracking the neutral X feature. In both cases, multiple resonances ($S_1$, $S_2$, $T_1$, and $T_b$), described in more detail below, are observed in addition to eCR. The internal transitions of the $X^-$-singlet $S_1$, $S_2$, and the $X^-$-triplet $T_1$ (see inset of Fig. 2 for the triplet states) occur at fields close to the previously observed and identified[4] transitions in quantum wells without growth interrupts. They correspond to ionizing transitions of an electron from a bound $X^-$-state associated with the lowest electron Landau level to the continuum of scattering states (hatched regions in the inset to Fig. 2) associated with the 2$^{nd}$ electron LL. In the high-field limit these can be thought of as CR-like transitions modified by the electron-electron and electron-hole Coulomb interactions.[3] In particular, transitions $S_1$ and $T_1$ correspond to CR-like transitions of the "second" electron in the $X^-$-complex in singlet and triplet states, respectively; the other two carriers of the complex (comprising X) remain in their respective ground states. The transitions labeled $S_2$ and $T_2$, on the other hand, correspond in this limit approximately to internal transitions of bound electron-hole pairs (X), from the ground singlet or triplet state to an excited p-like state with the "second" electron remaining in the lowest LL. The triplet transition $T_2$ occurs at much higher photon energies, i.e. much lower magnetic fields, outside of the region probed by the experiment and is therefore not observed. These are the only electric-dipole-allowed transitions in 2D QW systems for which both axial symmetry (the angular momentum selection rule) *and* magnetic translational invariance, are preserved.[3, 9]

In the high-field, strictly 2D limit for an ideal system, bound-to-bound triplet and singlet transitions, which dominate the spectra of negatively charged donors $D^-$ in QWs[6-8], are strictly forbidden for $X^-$. The dominant bound-to-bound triplet transition for $D^-$ lies at lower energy (higher magnetic field) than electron cyclotron resonance,[7] and calculations in the ideal high-field, strictly 2D limit[3, 9] show that the corresponding transition for X⁻ also lies at lower energy (higher field) than eCR. Random lateral potential energy fluctuations, whose range is greater than the effective Bohr radius of X⁻, can break the strict selection rules demanded by translational invariance and lead to non-zero oscillator strength for the bound-to-bound transitions. On the other hand, short range potential energy fluctuations (characteristic length < Bohr radius) are averaged out by the internal motion of the electrons and hole[21] and do not measurably affect the selection rules (see discussion below). Thus at the lattice temperature and excitation conditions of the experiment (effective carrier temperatures higher than the lattice temperature), we expect that most of the photo-excited electron-hole complexes behave as quasi-2D excitonic complexes, and that the monolayer fluctuations in well width perturb the ideal situation and weakly allow bound-to-bound transitions. For the constant photon energies of the ODR experiments the bound-to-bound triplet occurs on the high field side of eCR, while the bound-to-bound singlet transitions occur in the same field region as the bound-to-continuum features. A weak feature in the ODR spectra, labeled $T_b$ in Fig. 1(b), which occurs at a magnetic field *above* eCR (cf. inset of Fig. 2), is attributed to the bound-to-bound triplet transition.

Figure 2 shows the dependence of all observed transitions on FIR photon energy. The bold solid line is the calculated position of quasi-2D free electron CR for the average QW width with non-parabolicity included via a two-band $\bar{k} \cdot \bar{p}$ model.[22, 23] Numerical calculations of the positions of the onsets of the bound-to-continuum transitions for an ideal 20 nm GaAs/AlGaAs QW[4] are shown as crossed circles; the dashed lines are linear extrapolations of the calculated points. The field position of the bound-to-continuum transitions are difficult to determine experimentally for the two lowest FIR photon energies because they become increasingly broad and weak,[4] leading to the large error bars. This is also true for transition $T_b$, since the $X^-$-triplet becomes unbound at low-magnetic fields.[24]

To explore the effects of lateral confining potentials numerical calculations based on a 2D harmonic oscillator model were carried out. This model does not incorporate important features of the experimental situation for the wide well samples discussed here, namely the random and very weak nature of the potential energy landscape, which does not localize the excitons; however, it does contain the important ingredients of lateral potentials for narrow well samples for which excitons are confined. The general effect of parabolic lateral confinement is to shift all transition energies higher (fields lower) than eCR. This contrasts with the observations for the 14.1 nm well (excitons not localized), for which the additional feature occurs at <u>lower</u> photon energies (higher fields) than eCR. A shift to higher photon energies (lower fields) is indeed seen for the two narrowest wells of sample 2 (strongest lateral confining potentials);[11] these effects will be discussed in detail elsewhere.[25]

Figure 3(a) shows ODR field scans at several temperatures at $E_{FIR} = 10.43$ meV for sample 1[26], while Fig. 3(b) shows similar data for sample 2 (14.1 nm well width). Both samples show the same bound-to-continuum transitions, although the relative strengths are considerably different due to a strong dependence on electron density[4] and experimental conditions. More important, however, is the complete absence of any high-field shoulder above eCR for sample 1, which has no growth interrupts. In this case, the potential variations due to well-width fluctuations (or alloy fluctuations) are short range ($<<$ Bohr radius) and weak ($\approx 0.3$ meV); and averaging of these short range potential energy variations by the exciton complex's relative motion[21] leads to smooth, even weaker potential energy fluctuations that do not perturb magnetic translational symmetry sufficiently to allow an observable bound-to-bound triplet transition. In contrast, the intentionally introduced random well-width fluctuations in sample 2, produce much larger random potential energy variations of longer range than the Bohr radius of the complex, and thus relax the strict selection rule and permit the bound-to-bound transition $T_b$ to be observed. The latter is unobservable above 15 K, while the bound-to-continuum $X^-$-triplet transition $T_1$ is still seen. This is consistent with the excitonic complex losing sensitivity to the lateral potential fluctuations when the average thermal kinetic energy of the complex is larger than the magnitude of the potential energy fluctuations (0.75 meV). The $T_1$ transition, on the other hand, is a quasi-2D bound-to-continuum transition, and the initial (triplet ground) state lies slightly above the singlet ground states so that this transition initially grows slightly in strength as temperature is increased, and persists up to about 30 K. The loss of strength at higher temperature could be due to a number of factors including ionization of the initial triplet state and decreasing sensitivity of the ODR signal due to carrier heating as the lattice temperature increases.

In conclusion, ODR studies of negatively charged excitons have revealed a feature on the high field side of eCR in samples having QWs with intentional growth interrupts and relatively wide QWs (> 8.4 nm). Based on a detailed comparison with a sample grown without interrupts, extensive prior data on negatively charged donors, and calculations for ideal samples in the high field strictly 2D limit, we identify this feature as a triplet bound-to-bound transition.

This research was supported in part by NSF-DMR0203560 and NSF-IGERT DGE0114330. The work at CSUB was supported in part by a College Award of Cottrell Research Corporation. We are grateful to H. A. Nickel for use of the ODR data on sample 1.

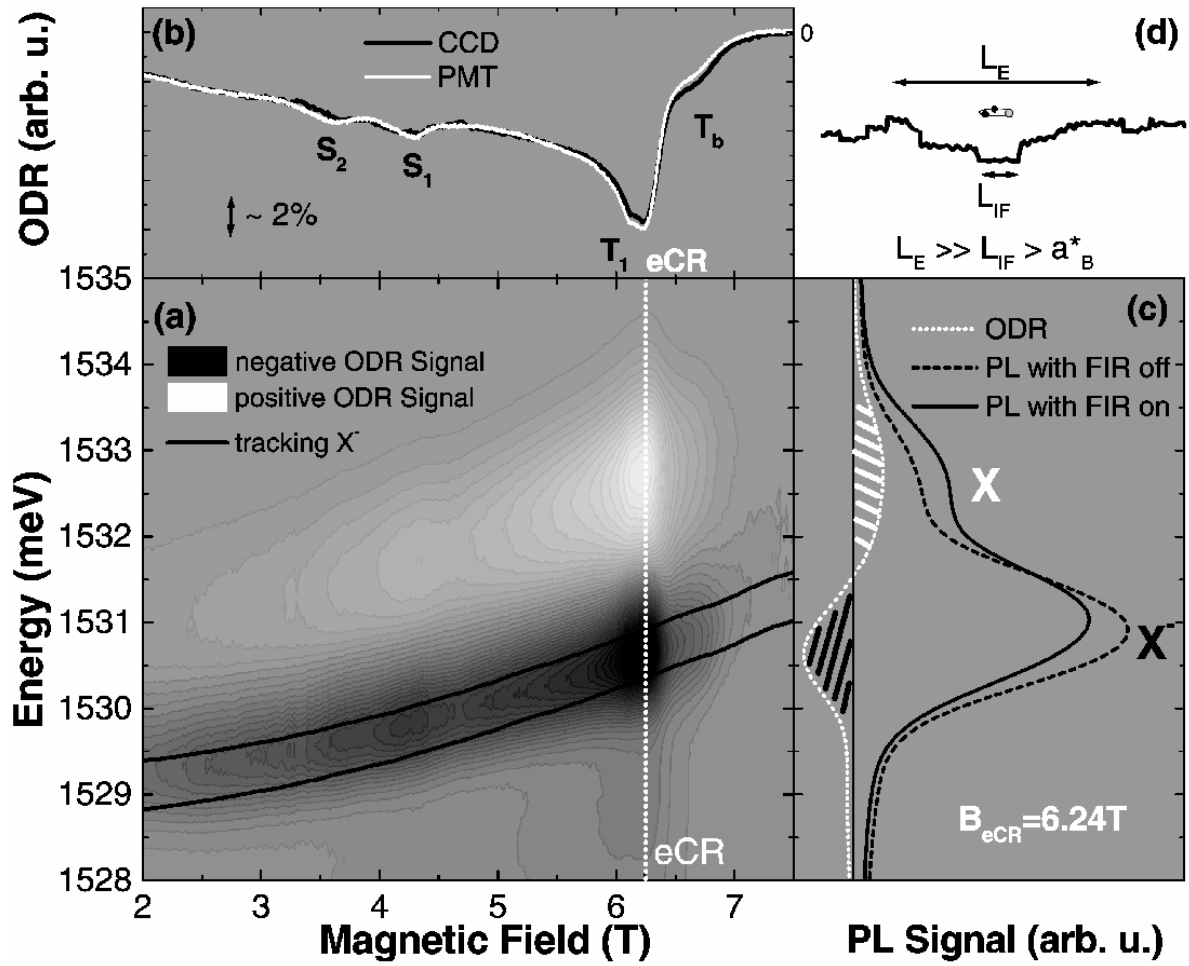

**Fig. 1:** *C. J. Meining et al.*, Phys. Rev. B

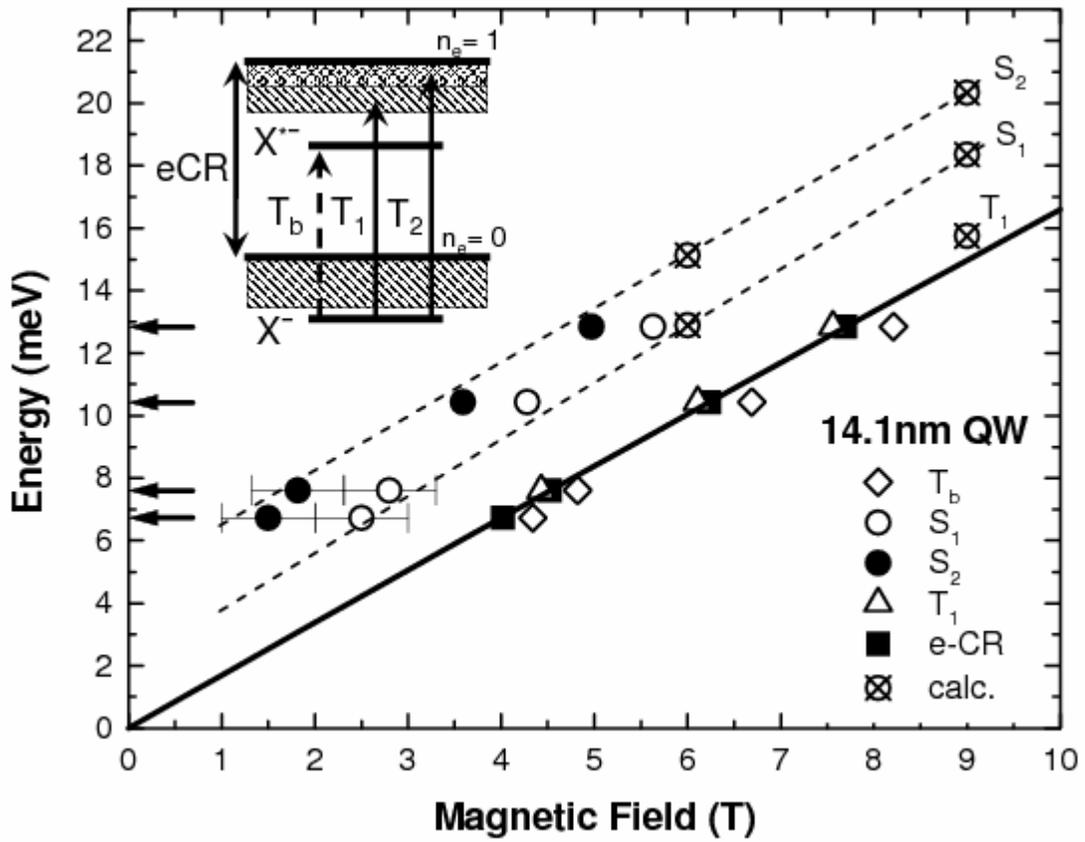

**Fig. 2:** *C. J. Meining et al.*, Phys. Rev. B

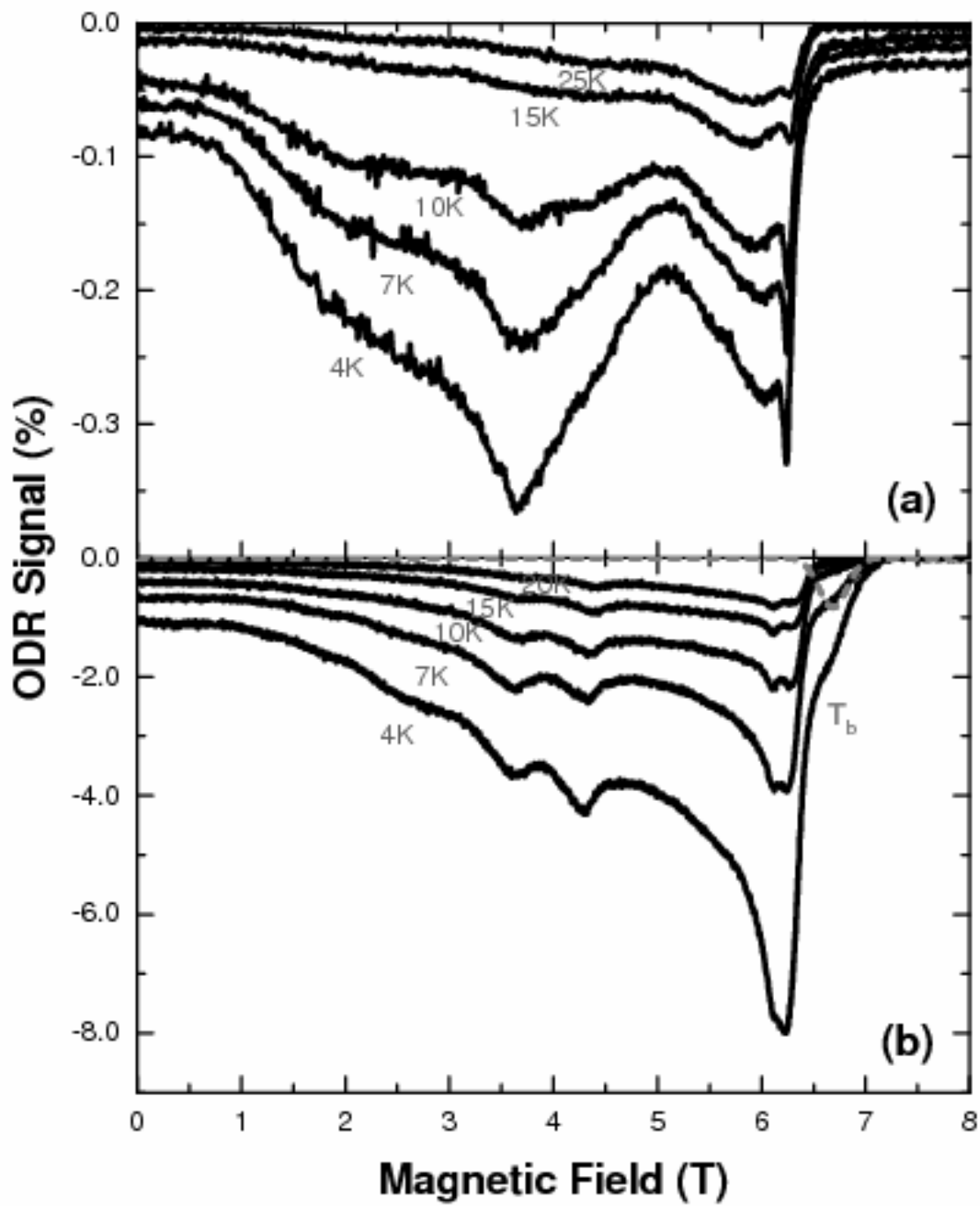

**Fig. 3:** *C. J. Meining et al.*, Phys. Rev. B

**Figure Captions:**

**Fig. 1:** ODR of a 14.1 nm wide QW at $E_{FIR}$ = 10.43 meV with PMT and CCD detection. **(a)** Gray scale contour plot of the CCD-ODR signal. Dark (bright) regions correspond to negative (positive) ODR signal. **(b)** ODR field scans obtained with CCD (thick black line) and PMT detection (thin white line). The pair of thick black lines in (a) indicate the bandpass used. Bound-to-continuum internal transitions of the X⁻-singlet ($S_1$, $S_2$), the X⁻-triplet ($T_1$), electron cyclotron resonance (eCR), and a bound-to-bound internal transition associated with the X⁻-triplet ($T_b$) are observed. **(c)** ODR energy scan (dotted white line) and CCD-PL with FIR laser on (solid black) and off (dashed black) at B = 6.24 T (the position of eCR is indicated by the vertical white dotted line in (a)). **(d)** Schematic of potential landscape: different length scales are indicated – $L_{IF}$ is the characteristic range of the interface fluctuations; $L_E$ is the range of the smooth electrostatic potential fluctuations due modulation doping; $a_B^*$ is the effective Bohr radius of the charged exciton.

**Fig. 2:** Magnetic field dependence of the observed transitions of sample 2 (14.1 nm QW; cf. Fig. 1(b)) at various fixed FIR photon energies: eCR (squares), bound-to-continuum internal transitions of the X⁻-singlet (circles), the X⁻-triplet (triangles), and the additional bound-to-bound, internal X⁻-triplet transition (diamonds). Solid black line: calculated eCR including non-parabolicity via a two-band model. Horizontal arrows indicate the FIR photon energies. The crossed circles mark numerical calculations at 6 T and 9 T for 20 nm QWs without lateral confinement; straight dashed lines are linear fits to the calculated points. Inset: schematic of the initial and final states of the X⁻-triplet and relevant FIR transitions associated with the two lowest electron LLs.

**Fig. 3:** Temperature study of the ODR signal for samples 1 **(a)** and 2 **(b)** for $E_{FIR}$ = 10.43 meV. In both panels the signal decreases with increasing temperature. Note the clear shoulder above eCR for sample 2; the dashed gray line shows the feature deconvolved from the background by fitting and subtracting the background contributions below eCR at 4.2 K.